\documentstyle[12pt,aasms4]{article}

\begin{document}

\noindent
{\it Conference Highlights}

\begin{center}


\title{\large \bf Stromlo Workshop on High-Velocity 
Clouds\footnote{Conference was held in Canberra, Australia in August 1998.
Proceedings will be edited by Brad K. Gibson \& Mary E. Putman
and published in the {\it ASP Conference Series}}} 

\end{center}

\medskip

While it may be somewhat unusual to claim that an area of research which
is only 35 years old is already undergoing a renaissance, this
statement is not that far from the truth, when one considers the field of
High-Velocity Clouds (HVCs).  A reflection of this assertion can be found by
surveying the literature of the past twelve months; what started with a
review in ARAA by Wakker \& van Woerden, culminated with the first
international workshop devoted solely to HVCs, held August 14th and 15th 1998
at Mount Stromlo Observatory.  In the interim, the field has been rife with new
developments, a few of which even received coverage
by the mainstream media, shedding much-needed positive light on this small, but
hearty, band of researchers.
The Stromlo Workshop on HVCs afforded the opportunity for each of
these researchers to bring us up to speed on this rapidly evolving field.
The summary which follows cannot do justice to all the workshop participants,
and at some level is almost certainly biased by the present authors'
prejudices, for which we apologize in advance.

Key contributions since the 1997 Wakker \& van Woerden review include - 
(a) the first determinations of a true
metallicity for an HVC (i.e. $\sim 0.25$\,Z$_\odot$ for HVC~287.5+22.5+240,
from Lu et~al., and $\sim 0.1$\,Z$_\odot$ for Complex~C, from Wakker et~al.);
(b) the recognition of a leading arm counter-feature to the Magellanic Stream
by Putman et~al., indicative of a tidal origin; 
(c) the first determination of a distance \it bracket \rm for an HVC (i.e.
$d=4\rightarrow 11$\,kpc for Complex~A, from van Woerden et~al.);
(d) the burgeoning cottage
industry of emission line (primarily H$\alpha$) observations of HVCs, led
most noticeably by Tufte et~al. and Bland-Hawthorn et~al; and
(e) the proposal by Leo Blitz et~al. that HVCs are intergalactic remnants of 
the formation of the Local Group.

While all-sky \ion{H}{1} surveys for HVCs have existed since the late-1980s,
a major step forward in ensuring uniformity in spatial and velocity resolution
can be expected in the coming year with the release of Villa Elisa Southern
Sky Survey (VESSS;
Morras et~al.), a direct analog to the Leiden-Dwingeloo Survey (LDS) in
the north.  Higher angular resolution surveys, on a sub-Nyquist sampled spatial
grid, will ultimately be
available once the Parkes Multibeam HVC Survey (Putman et~al.) completes its
scans of the sky from $\delta=-90^\circ\rightarrow +25^\circ$.
A crucial first step will be
the compilation of a new catalog of anomalous-velocity \ion{H}{1} gas, from the
LDS and VESSS, with far more stringent selection criteria than adopted
previously.  Cross-correlating with the Parkes Multibeam HVC Survey will shed
light on any residual compact, isolated HVCs missed by the somewhat
coarser LDS and
VESSS.  Putman et~al. also report the initiation of a long-term high-resolution
study of a number of southern HVCs with the Compact Aray at Narrabri; a primary
use for this data will be in deriving accurate \ion{H}{1} column densities to
background stellar and extragalactic probes for follow-up metallicity work with
HST and the Far-Ultraviolet Spectroscopic Explorer (FUSE).  The FUSE mission
was described nicely by Ken Sembach.

The first determination of absolute
metallicity, for two different HVCs,
were presented by Bart Wakker - one (HVC~287.5+22.5+240)
is clearly highly enriched and consistent with being
tidally-disrupted from the Magellanic Clouds, while the other (part of
Complex~C) is closer to $\sim 0.1$\,Z$_\odot$.  
While suggestive, it is perhaps
premature to insist that \it all \rm HVCs therefore have strongly sub-solar
metallicities.  
Low (but perhaps not primordial) abundances are a natural
prediction of the model described by Blitz et~al., in which many HVCs are
presumed to be remnants from the formation of the Local Group.
While controversial, Blitz
et~al. have succeeded in focusing a great deal of attention back upon HVCs, and
their model does have the beauty of making very testable predictions.

One prediction that is
currently being put to the test, and one emphasized by Leo Blitz in his talk,
is that HVCs should not
show evidence for any H$\alpha$ emission.  
We heard from Joss 
Bland-Hawthorn and his TAURUS-II team, as well as Steve Tufte from
the WHAM team, both of whom are overseeing ambitious programs of HVC
emission-line studies using Fabry-Perot interferometers.  Between the two
teams, emission measures of several elements have been measured for three major
Complexes (WHAM) and one isolated large HVC (TAURUS-II).  Both teams are
currently targeting compact, isolated HVCs, as a direct (simultaneous)
test of both (a) the
Blitz et~al. model, and (b) the Bland-Hawthorn \& Maloney Galactic halo
ionizing radiation field model.

Discussion of the WHAM and TAURUS-II programs naturally led into what was
nearly a full day devoted to hot gas and the Galactic halo, kicked off
by Ulrich Mebold's review, followed by
key contributions from Laura Danly, Peter 
Kalberla and Don York.  
The highly-ionized ``\ion{C}{4}-HVCs'', seen along the line-of-sight
to PKS~2155-304 and Mrk~509, were then 
described by Ken Sembach;  the implied low thermal pressures involved
are suggestive of an intergalactic location.
Finally, Bob Benjamin and Jacques Lepine left us speculating about the
consequences of HVCs impacting upon the Galactic disk, and their role in
driving star formation in the solar neighborhood.

Hugo van Woerden's review of the techniques and subtleties
of HVC distance determination, helped focus attention on the existing
shortcomings in the field - i.e., the lack of suitable stellar
probes, of known distance, aligned (both foreground \it and \rm background)
along the lines-of-sight to suitable HVCs.  This dearth of stellar probes
cropped up repeatedly during the Workshop, so much so that a small working
group was convened by Bart Wakker in order to clarify exactly what steps can
be taken over the coming few years to rectify the situation.  Programs aimed at
uncovering numerous blue stars in the halo were described by Tim Beers, Silvia
Rossi, and Jeff Pier, and are \it eagerly \rm anticipated by the HVC community
at large.  Hugo's talk concluded with their recent results on determining the
distance to Complex~A; for the first time, both an upper \it and \rm lower
limit for an HVC exists, the distance bracket being $4\rightarrow 11$ kpc.

Attempting to give the wild speculations of many of the above 
observers some grounding in
theory was the thankless task assigned Joel Bregman, Miguel de Avillez and
Mark Walker.

That most favorite of HVCs, the Magellanic Stream, had a session set aside for
itself.  Mary Putman presented what is, to date, the best direct evidence for
a tidal origin to the Stream, a counter-feature stretching
more than 25$^\circ$, leading the Stream and Magellanic Clouds, supporting the
indirect evidence (presented by Bart Wakker) regarding the earlier
Cloud-like
metallicity determination by Limin Lu et~al. for HVC~287.5+22.5+240.  
Lance Gardiner showed new n-body simulations which built upon his already
impressive tidal models, but with significantly improved recovery of the
Leading Arm's spatial and velocity structure, and Raymond Haynes described an
impending survey of the
Clouds and Leading Arm region which will use the new narrow-band upgrade
on the Parkes Multibeam facility.

Closer to home, both Mike Gladders and Bryan Penprase reminded us that the
study of HVCs should not be done with blinders on; understanding the origin,
properties, and environment of
Intermediate-Velocity Clouds (IVCs) and high-latitude molecular clouds, may
be inextricably linked with that of HVCs.

As noted at the outset, 
doing justice to all the contributors in this Highlights paper proved to be
an impossible task.  We encourage the
reader to seek out a copy of the Proceedings, in order to get more 
complete (and impartial?) coverage of the Workshop.

On a personal note, the Organizing Committee (Brad Gibson and Mary Putman)
wish to thank each and every one of the participants for their
contributions.  If the measure of a meeting's success is the degree of animated
post-talk debate, then it is clear that the Stromlo Workshop on High-Velocity
Clouds exceeded even our own fairly lofty ambitions.

\medskip

\noindent
{\it Brad K. Gibson$^{\rm 1}$ \& Bart P. Wakker$^{\rm 2}$

\noindent
{\it $^{\rm 1}$University of Colorado}

\vskip-6.0mm
\noindent
{\it $^{\rm 2}$University of Wisconsin}

\end{document}